\documentclass{nature}

\usepackage{url}
\usepackage{amsmath}
\usepackage{amssymb}
\usepackage{graphicx}
\usepackage{bm}

\title{Neutron Star Observations Challenge a Large Colour-Superconducting Gap in Dense Quark Matter}

\begin{document}

\maketitle
\author{Shao-Peng~Tang$^{1}$, Yong-Jia~Huang$^{1,2}$, Yi-Zhong~Fan$^{1,3,\ast}$.}

\begin{affiliations}
\small
\item{Key Laboratory of Dark Matter and Space Astronomy, Purple Mountain Observatory, Chinese Academy of Sciences, Nanjing 210033, China}
\item{RIKEN Center for Interdisciplinary Theoretical and Mathematical Sciences (iTHEMS), RIKEN, Wako 351-0198, Japan}
\item{School of Astronomy and Space Science, University of Science and Technology of China, Hefei, Anhui 230026, China}\\
$^\ast$Corresponding author. Email: yzfan@pmo.ac.cn
\end{affiliations}

\hfill

\begin{abstract}
At asymptotically high density, quantum chromodynamics (QCD) predicts that quark matter becomes a colour superconductor in the colour-flavour-locked (CFL) phase, yet, away from the asymptotic regime, the magnitude of the pairing gap is uncertain.
Here we combine multi-messenger neutron star observations with perturbative QCD and chiral effective field theory inputs in a Bayesian inference built on a flexible Gaussian-process--neural-network representation of the equation of state (EOS).
By matching the EOS at a baryon chemical potential of 2.6 GeV to the perturbative QCD prediction supplemented by the next-to-leading-order CFL contribution, we infer a gap of $\Delta^{*}_{\rm CFL}=34^{+32}_{-28}$ MeV and a 95\% credible upper limit of about 66 MeV, which is a factor of two tighter than previous bounds and at the lower edge of most microscopic model predictions.
We further place the first data-driven constraint on the unknown high-order constant of the perturbative QCD pressure, i.e., $c_0=-21^{+9}_{-8}$.
Our results indicate that colour-superconducting pairing makes only a subdominant contribution to dense matter pressure, tightening the connection between neutron star data and the QCD phase diagram.
\end{abstract}


Together with high-density perturbative QCD (pQCD) and low-density chiral effective field theory ($\chi$EFT), the recent multi-messenger observations of neutron stars (NSs), including precise mass determinations of heavy pulsars, NICER mass$-$radius measurements, and tidal deformability constraints from GW170817, have substantially narrowed the range of admissible dense-matter equations of state (EOSs) \cite{2020NatPh..16..907A, 2022ApJ...939L..34A, 2023ApJ...950..107G, 2023SciBu..68..913H, 2023ApJ...949...11J, 2023NatCo..14.8451A,2023PhRvD.108i4014B, 2024PhRvD.109d3052F, 2025PhRvD.112h3009T}.
Robust features have emerged, such as a peak in the sound-speed squared $c_{\rm s}^2$ above the conformal value $1/3$ at densities $\sim (3-5)\,n_{\rm s}$ and a softening of the EOS in the highest-density region reached by stable NSs \cite{2023ApJ...950..107G, 2023SciBu..68..913H, 2023NatCo..14.8451A, 2026ApJ..1002...40G, 2026arXiv260508584H}.
Equally importantly, the recent completion of the N$^3$LO pQCD pressure (modulo a single unknown constant $c_0$) \cite{2023PhRvL.131r1902G} together with the integral causality/stability/thermodynamic-consistency constraints \cite{2022PhRvL.128t2701K} propagate information from very high baryon chemical potential down to NS densities, further sharpening the EOS posterior.
Because the CFL color-superconducting pairing modifies the pQCD boundary condition through a contribution $p_{\rm CFL}\sim \Delta_{\rm CFL}^2\mu_B^2/(3\pi^2)$ \cite{1998PhLB..422..247A, 1999NuPhB.537..443A, 2008RvMP...80.1455A}, astrophysical EOS constraints reciprocally bound the magnitude of the CFL gap.
Color-flavor locking is the natural ground state of three-flavor quark matter at asymptotically high density \cite{1999NuPhB.537..443A}; pinning down $\Delta_{\rm CFL}$ in the regime $\mu_B\gtrsim 2$ GeV bears directly on the phase structure of QCD and on transport, cooling, and r-mode damping in compact stars \cite{2008RvMP...80.1455A}.

While weak-coupling QCD reliably predicts the gap only at asymptotically high density, where $\Delta_{\rm CFL}\propto\mu_q\,g^{-5}\exp[-3\pi^2/(\sqrt{2}\,g)]$ with $\mu_q=\mu_B/3$ \cite{1999PhRvD..59i4019S}, its value in the more moderate density regime is not quantitatively known and must be estimated from models.
Such calculations span a broad range, $\Delta_{\rm CFL}\sim 20-250$ MeV and most typically $\sim 50-150$ MeV \cite{1998PhLB..422..247A, 1998PhRvL..81...53R, 1999NuPhB.537..443A, 1999NuPhB.538..215B, 1999PhRvD..60a6004C, 2008RvMP...80.1455A, 2018RPPh...81e6902B, 2020PhRvL.125n2502L}, with one recent estimate as large as $\sim 300$ MeV \cite{2022PhRvD.105c6003B}.
Two recent works have used model-agnostic Bayesian EOS inference to place an upper bound on the CFL gap. Ref.~\cite{2024PhRvL.132z2701K} first showed at leading order (LO) that the integral pQCD constraints already exclude large gaps, finding $\Delta_{\rm CFL}^{*}(2.6\,{\rm GeV})\lesssim 216$ MeV in a ``reasonable'' scenario.
Subsequently, ref.~\cite{2025PhRvL.135u1901G} computed the NLO corrections to the CFL pressure under NS conditions, showing that the NLO term $\gamma_1=4-4\bar m_s^2/3+40.9\,\alpha_s$ enhances the CFL pressure substantially.
Folding the NLO formula into the same Bayesian framework and allowing for a power-law gap $\Delta_{\rm CFL}\propto\mu_B^{\sigma}$, they obtained $\Delta_{\rm CFL}^{*}\lesssim 140$ MeV at 95\% credibility.
While encouraging, these analyses are conservative in that they do not exploit detailed structural information of the EOS in the density window $\sim n_{\rm TOV}-40\,n_{\rm s}$ which sits between the NS interior and the pQCD region.

Recent non-parametric analyses have shown that global thermodynamic consistency can effectively narrow down the EOS behavior in this intermediate-density regime \cite{2026arXiv260508584H}.
The present work is motivated by the prospect of using this intermediate window to strengthen the constraint.
The GP-based nonparametric frameworks employed in these analyses, however, are typically explored by drawing a large prior ensemble of EOS realizations and reweighting them by the likelihood, with the high-density pQCD information imposed a posteriori as an integral consistency cut \cite{2023ApJ...950..107G, 2024PhRvL.132z2701K, 2025PhRvD.112h3009T, 2026arXiv260508584H}.
While efficient when the high-density boundary is fixed, this strategy becomes prohibitive once the boundary condition itself depends on several pQCD and CFL parameters ($c_0$, $\Delta_{\rm CFL}^{*}$, $\sigma$): mapping their joint posterior would then require a dense grid scan over a high-dimensional space, which is computationally expensive and prone to under-sampling the physically relevant region, so that convergence to the true posterior is not guaranteed.
This motivates an EOS representation that is simultaneously flexible yet not so high-dimensional as to become intractable for nested sampling, in which the dense-matter EOS and the pQCD$+$CFL matching parameters are sampled jointly and the Bayesian evidence is obtained directly.

To this end we develop a hybrid, flexible parameterization that combines a fine multivariate-Gaussian representation of $c_{\rm s}^2$ in $\phi\equiv-\ln\!\left(1/c_{\rm s}^2-1\right)$ space within and slightly above NS densities, with a boundary-constrained feed-forward neural-network (NN) representation at higher density up to the pQCD matching point.
The GP is implemented through a non-centered (whitening) transform, with the correlation length $\ell_{\rm GP}$, variance $\sigma^2_{\rm GP}$, asymptotic sound speed $\bar c_{\rm s}^2$, and the prior-mean transition density $n_{\rm end}$ treated as sampled hyperparameters.
This allows the data to inform the smoothness and amplitude of the intermediate-density EOS rather than fixing them by hand.
The NN bridge encodes multi-scale structure across the wide region $[8\,n_{\rm s},n_{\rm match}]$ with only a single hidden layer of $10$ neurons under Xavier-scaled, shape-regularizing priors (non-negative output weights and monotonically ordered hidden biases), while its baseline-plus-envelope construction enforces the boundary values exactly.
The model thereby retains the flexibility of nonparametric approaches \cite{2021ApJ...919...11H, 2019PhRvD..99h4049L, 2020PhRvD.101f3007E, 2020PhRvD.101l3007L} while keeping the dimensionality at a level tractable for nested sampling, yielding sharper marginal posteriors and reliable evidence estimates.
With this representation, the boundary at $\mu_B=2.6$ GeV is matched directly to the pQCD$+$CFL prediction, using the N$^3$LO pQCD pressure \cite{2023PhRvL.131r1902G} together with the NLO gap correction \cite{2025PhRvL.135u1901G}.
We obtain $\Delta_{\rm CFL}^{*}=34^{+32}_{-28}\;{\rm MeV}$, corresponding to a factor-of-two reduction in the 95\% credible upper limit relative to ref.~\cite{2025PhRvL.135u1901G}, with a posterior that peaks at a nonzero value.
The same analysis yields a posterior on the N$^3$LO unknown $c_0=-21^{+9}_{-8}$, providing the first data-driven constraint on this constant.
To gauge the impact of color superconductivity we also repeat the entire inference without the CFL term, finding consistent EOS posteriors throughout.

\textbf{\emph{Results}}

We performed a joint Bayesian inference with the hybrid GP--NN EOS parameterization and astrophysical likelihood, matching the high-density boundary at $\mu_B=2.6$ GeV to the N$^3$LO pQCD pressure and density augmented by the NLO color-flavor locked contribution.
Below we present the resulting constraints on the CFL pairing gap, the N$^3$LO constant $c_0$, and the inferred dense-matter EOS, and we contrast the fiducial analysis with a control run that omits the CFL contribution.

\paragraph{The CFL pairing gap.}

\begin{figure}[!h]
\centering
\includegraphics[width=0.618\linewidth]{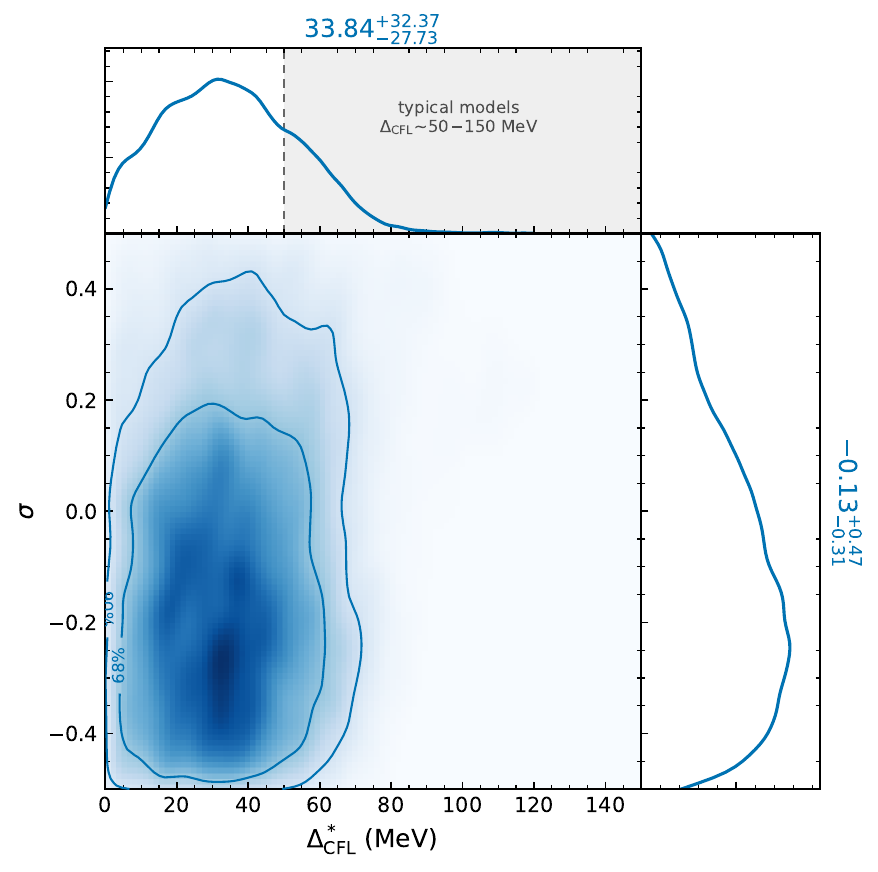}
\caption{{\bf Joint posterior of the CFL gap magnitude $\Delta_{\rm CFL}^{*}=\Delta_{\rm CFL}(\mu_B=2.6~{\rm GeV})$ and the running exponent $\sigma$.}
Contours mark the 68\% and 90\% credible regions.
The top panel shows the marginalized posterior of $\Delta_{\rm CFL}^{*}$; the grey band indicates the typical range $\Delta_{\rm CFL}\sim 50\!-\!150$ MeV from most microscopic models \cite{2008RvMP...80.1455A}, assuming a density-independent gap for comparison.}
\label{fig:corner_gap}
\end{figure}

Figure~\ref{fig:corner_gap} shows the joint posterior of the CFL gap magnitude $\Delta_{\rm CFL}^{*}=\Delta_{\rm CFL}(\mu_B=2.6~{\rm GeV})$ and the running exponent $\sigma$.
The marginalized one-dimensional posterior on $\Delta_{\rm CFL}^{*}$ peaks at a nonzero value,
\begin{equation}
\Delta_{\rm CFL}^{*}=33.84^{+32.37}_{-27.73}~{\rm MeV}\quad(90\%~{\rm CI}),
\end{equation}
with a 95\% credible upper limit of $\Delta_{\rm CFL}^{*}\lesssim 66$ MeV, which is a factor of $\sim 2$ tighter than the $\lesssim 140$ MeV bound of ref.~\cite{2025PhRvL.135u1901G}.
For a nearly density-independent gap this limit falls at the lower edge of the range spanned by microscopic predictions ($\Delta_{\rm CFL}\sim 50-150$ MeV in most models) \cite{2008RvMP...80.1455A}, so that current data already disfavor all but the smallest theoretically anticipated gaps.
Notably, the early NJL-type estimate of Alford et al.~\cite{1999NuPhB.537..443A}, whose range extends down to 10 MeV, remains compatible with our posterior.
The improvement over previous model-agnostic analyses stems from two ingredients: the exploitation of the structurally informative intermediate-density window through the NN bridge, and the use of nested sampling to map the joint $(\Delta_{\rm CFL}^{*},\sigma,c_0,X)$ posterior directly rather than reweighting a fixed prior ensemble.
The running exponent is only weakly informed, $\sigma=-0.13^{+0.47}_{-0.31}$, with a broad posterior consistent with a density-independent (constant) gap.

\paragraph{Matching at the pQCD boundary.}

\begin{figure}[!h]
\centering
\includegraphics[width=0.618\linewidth]{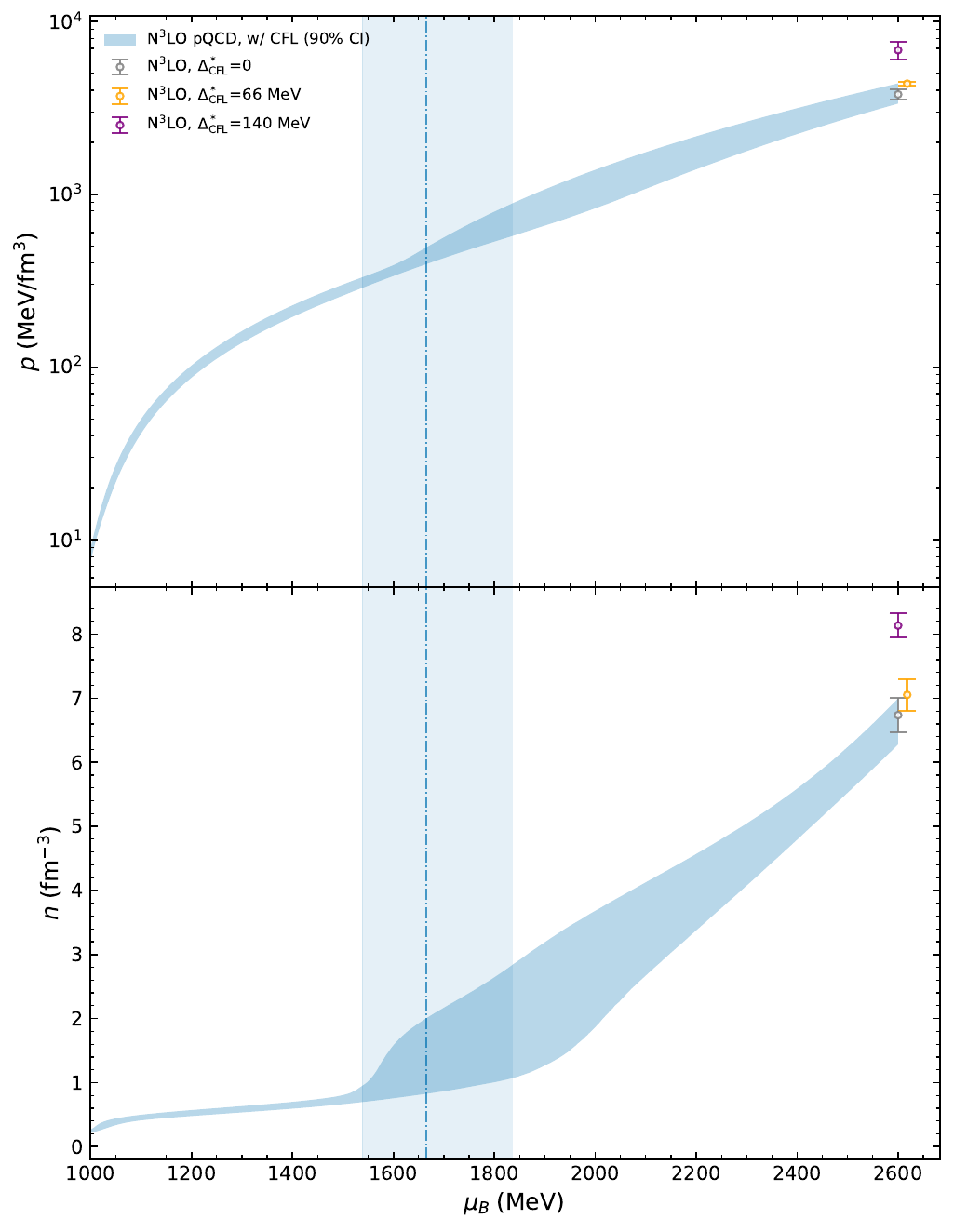}
\caption{{\bf Top: posterior band (90\% CI) of the pressure $p(\mu_B)$ from the N$^3$LO pQCD$+$CFL inference, compared with the pQCD$+$CFL boundary prediction at $\mu_B=2.6$ GeV for $\Delta_{\rm CFL}^{*}=0,\,66,\,140$ MeV.
Bottom: the same comparison for the baryon density $n(\mu_B)$.}
The reference markers are computed with $\sigma=0$ and $c_0$ fixed to its posterior median, while the error bars indicate the variation induced by marginalizing $X$ over its log-uniform prior $[0.5,2]$.
Larger gaps push the boundary $(p,n)$ upward, increasing the tension with the EOS.
The vertical shaded band indicates the 90\% CI of the central chemical potential for the maximum-mass ($M_{\rm TOV}$) configuration.}
\label{fig:mu_p_n}
\end{figure}

Figure~\ref{fig:mu_p_n} illustrates how the constraint operates by comparing the posterior bands of the pressure $p(\mu_B)$ (top) and baryon density $n(\mu_B)$ (bottom) with the pQCD$+$CFL boundary prediction at $\mu_B=2.6$ GeV for representative gap values $\Delta_{\rm CFL}^{*}=0,\,66,\,140$ MeV.
For a clean comparison, these reference points are evaluated with the running exponent fixed to $\sigma=0$ (constant gap) and the N$^3$LO constant set to its posterior median $c_0\simeq-21$, while the renormalization-scale ratio $X$ is marginalized over its log-uniform prior $[0.5,2]$; the resulting spread in $X$ is shown as the error bars on each marker, corresponding to the minimum and maximum allowed values.
Because the NLO CFL term enters as $p_{\rm CFL}^{\rm NLO}\propto\gamma_1\bar\Delta_{\rm CFL}^2\mu_B^4$ with $\gamma_1=4-\tfrac{4}{3}\bar m_{\rm s}^2+40.9\,\alpha_{\rm s}$, a larger gap shifts both the boundary pressure and density upward, increasing the tension with the EOS extrapolated from NS densities.
The $\Delta_{\rm CFL}^{*}=140$ MeV point lies well above the 90\% posterior band in both pressure and density, whereas $\Delta_{\rm CFL}^{*}\lesssim 66$ MeV remains compatible with the density band but sits marginally at the upper edge of the pressure band.
The NLO corrections amplify the leading-order shift substantially, which both sharpens the upper limit and may account for the mild preference for a small but finite gap: a modest nonzero $\Delta_{\rm CFL}^{*}$ raises the boundary pressure just enough to improve the joint matching to the inferred intermediate-density EOS.

\paragraph{The N$^3$LO constant $c_0$.}

\begin{figure}[!h]
\centering
\includegraphics[width=0.618\linewidth]{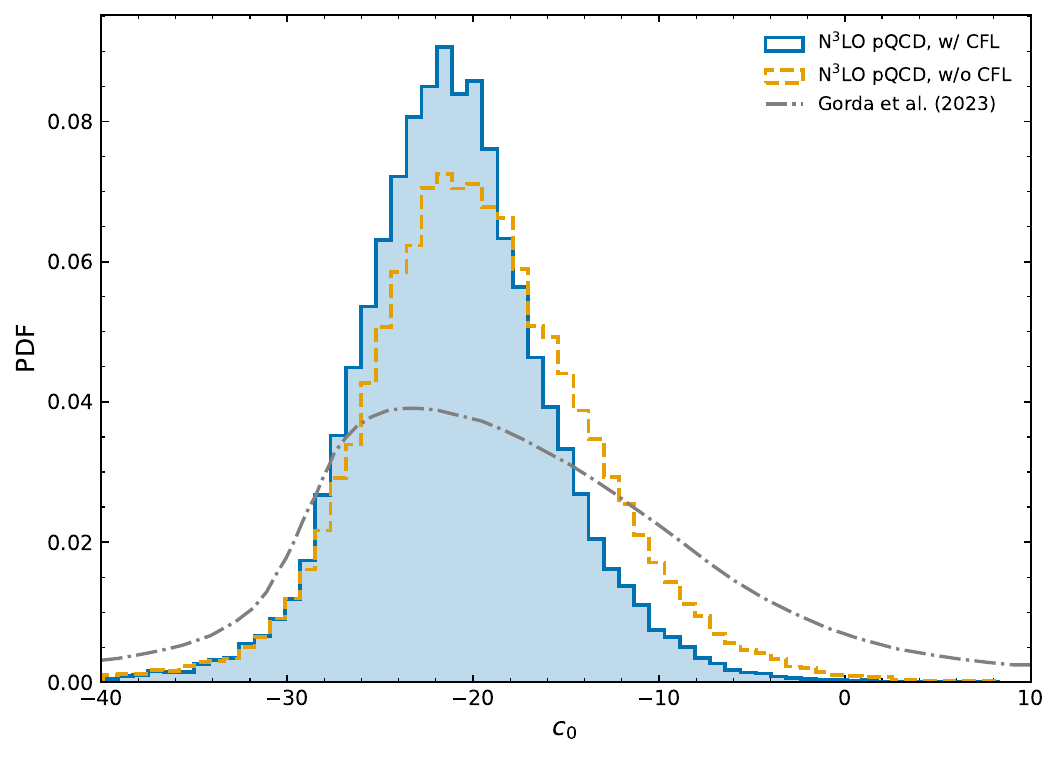}
\caption{\bf Marginal posteriors of the N$^3$LO constant $c_0$ for the inferences with (w/ CFL) and without (w/o CFL) the color-superconducting contribution, compared with the prior estimate from the convergence analysis of Gorda et al.~(2023) \cite{2023PhRvL.131r1902G}.}
\label{fig:c0}
\end{figure}

The previously unknown N$^3$LO constant $c_0$ is constrained simultaneously.
Figure~\ref{fig:c0} shows its marginal posterior, together with the convergence-analysis estimate of ref.~\cite{2023PhRvL.131r1902G}.
We obtain
\begin{equation}
c_0 = -21.2^{+8.8}_{-7.5}\quad(90\%~{\rm CI}),
\end{equation}
i.e., current astrophysical data already select a well-defined region of the N$^3$LO parameter space.
The constraint is driven by the requirement of a thermodynamically consistent connection between NS densities and $\mu_B=2.6$ GeV, and the corner plot (Supplementary Information) shows that $c_0$ is only mildly correlated with $\Delta_{\rm CFL}^{*}$.
This is, to our knowledge, the first data-driven posterior on $c_0$ from a joint NS$+$pQCD inference; it lies comfortably within the range spanned by the theoretical prior, demonstrating that the empirical determination is compatible with the first-principles convergence estimate.
We have also tested the no-CFL case, which yields $c_0=-19.9^{+10.7}_{-8.9}$; this slight shift toward less negative values is expected because both a larger $c_0$ and a nonzero CFL gap raise the boundary pressure, and the two effects partially compensate when matching to the EOS.

\paragraph{The inferred equation of state.}

\begin{figure}[!h]
\centering
\includegraphics[width=0.618\linewidth]{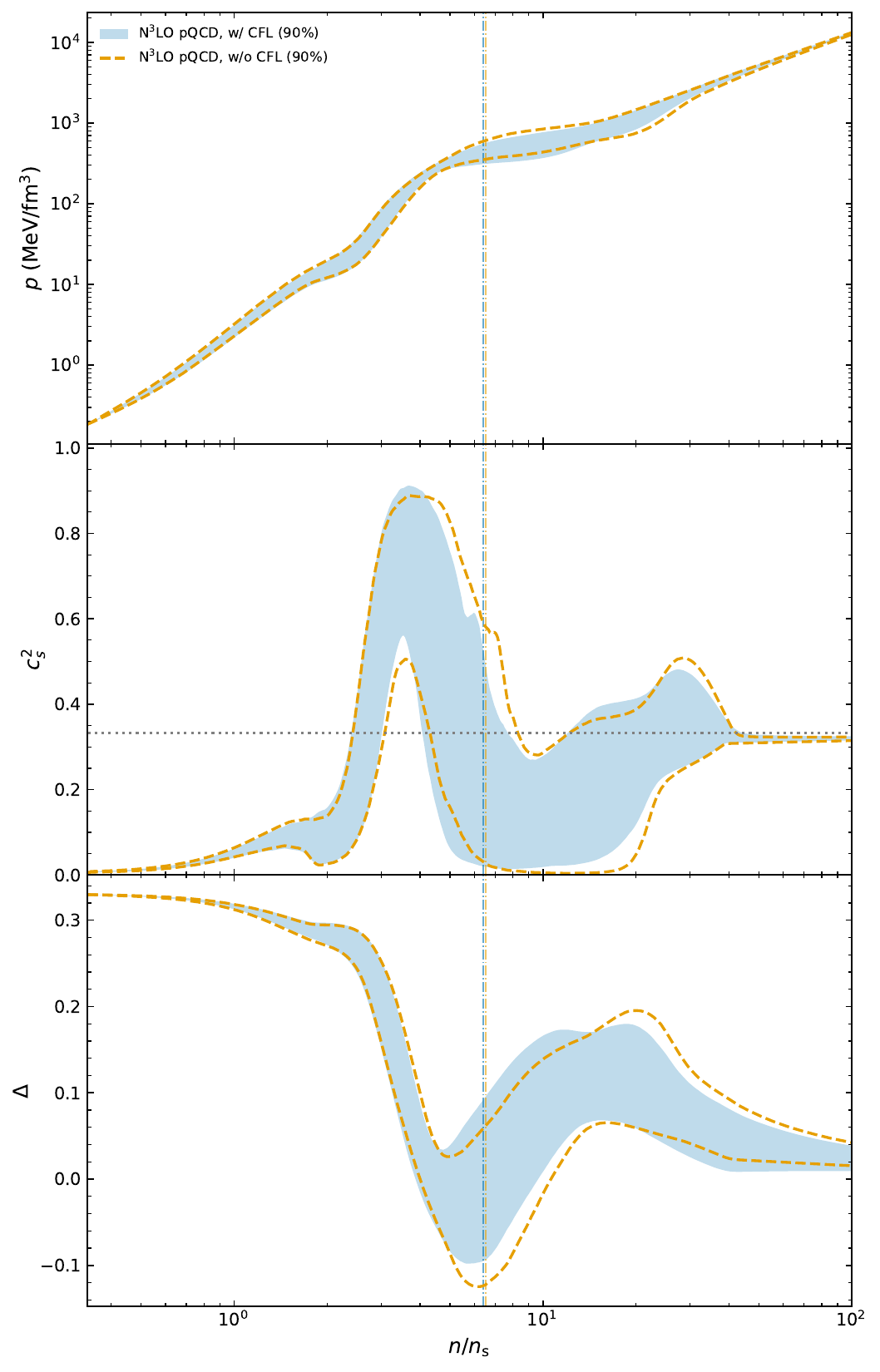}
\caption{{\bf Posterior 90\% bands for the pressure (top), the squared sound speed $c_{\rm s}^2$ (middle), and the normalized trace anomaly $\Delta=1/3-p/\epsilon$ (bottom) \cite{2022PhRvL.129y2702F} as functions of $n/n_{\rm s}$, for the inferences with (w/ CFL) and without (w/o CFL) the color-superconducting contribution.}
The vertical lines indicate the median central density of the maximum-mass ($M_{\rm TOV}$) configuration.}
\label{fig:eos}
\end{figure}

Figure~\ref{fig:eos} presents the posterior 90\% bands for the pressure $p(n)$, the squared sound speed $c_{\rm s}^2(n)$, and the normalized trace anomaly $\Delta=1/3-p/\epsilon$ over the full density range.
We recover the now-canonical features of recent EOS inferences: a $c_{\rm s}^2$ peak that exceeds the conformal value $1/3$ near $\sim(3-5)\,n_{\rm s}$, a subsequent softening around the TOV density, and a slow, non-monotonic approach toward the conformal limit from below, accompanied by a trace anomaly that turns over and tends toward zero at high density.
Because the inferred gap is modest ($\Delta_{\mathrm{CFL}}^{*} \lesssim 66~\mathrm{MeV}$), the pairing pressure $p_{\mathrm{CFL}}^{\mathrm{NLO}} \propto \bar{\Delta}_{\mathrm{CFL}}^{2}$ constitutes only a subdominant correction to the N$^3$LO pQCD pressure at $\mu_B = 2.6~\mathrm{GeV}$.
The high-$\mu_B$ boundary thus constrains $\Delta_{\mathrm{CFL}}^{*}$ predominantly through the joint $(p,n)$ matching while leaving the reconstructed $c_{\mathrm{s}}^{2}(n)$, $p(n)$, and trace anomaly essentially unchanged.
Indeed, the two bands in Fig.~\ref{fig:eos} (with and without the CFL terms) are nearly indistinguishable across the entire density range, directly confirming that color-superconducting pairing is a sub-dominant correction at the matching point.
The NN bridge interpolates smoothly across the wide intermediate window without introducing spurious oscillations, while the right anchor enforces consistency with the asymptotic pQCD$+$CFL behavior.

The macroscopic NS properties derived from these EOS posteriors are summarized in the Supplementary Information:
$M_{\rm TOV}=2.16^{+0.10}_{-0.09}\,M_\odot$, $R_{1.4}=11.49^{+0.56}_{-0.43}$ km, $R_{2.0}=11.37^{+0.56}_{-0.49}$ km, and $\Lambda_{1.4}=250^{+98}_{-56}$, with a central density at the TOV configuration of $n_{\rm TOV}=6.40^{+0.70}_{-0.64}\,n_{\rm s}$.
The corresponding no-CFL run yields statistically indistinguishable values (e.g., \ $M_{\rm TOV}$ and $R_{1.4}$ shift by well below their $1\sigma$ uncertainties), underscoring the robustness of the reconstructed EOS against the inclusion of the pairing term.
Notably, the inferred value of $M_{\rm TOV}$ is broadly consistent with the estimate based on the multimessenger observations of GW170817/GRB 170817A/AT2017gfo \cite{2020PhRvD.101f3029S, 2020ApJ...904..119F}, suggesting that the merger remnant was a temporary neutron star.

\textbf{\emph{Discussion}}

We have carried out a joint Bayesian inference of the dense-matter EOS and the high-density QCD boundary using a hybrid Gaussian-process--neural-network representation of the sound speed, matched at $\mu_B=2.6$ GeV to the N$^3$LO pQCD pressure and density including the NLO color-flavor locked contribution.
By sampling the EOS, the pQCD constant $c_0$, the renormalization scale $X$, and the CFL gap parameters ($\Delta_{\rm CFL}^{*}$,$\sigma$) jointly with nested sampling, we obtain three principal results.

First, the inference yields a $95\%$ credible upper limit $\Delta_{\rm CFL}^{*}\lesssim 66$ MeV at $\mu_B=2.6$ GeV, a factor of $\sim 2$ tighter than the previous model-agnostic bound of ref.~\cite{2025PhRvL.135u1901G}, with a one-dimensional posterior that peaks at a nonzero value $\Delta_{\rm CFL}^{*}=33.84^{+32.37}_{-27.73}$ MeV.
The tightening relative to ref.~\cite{2025PhRvL.135u1901G} has three complementary origins:
(i) the inclusion of structurally informative intermediate-density EOS and the joint $(p,n)$ matching;
(ii) the use of more recent and more numerous NICER/$M_{\mathrm{max}}$ inputs; and
(iii) a compact parameterization sampled jointly with the boundary parameters by nested sampling, yielding reliable, well-converged marginals and evidences.

Second, the analysis delivers the first data-driven posterior on the N$^3$LO constant, $c_0=-21.2^{+8.8}_{-7.5}$, consistent with the convergence-analysis estimate of ref.~\cite{2023PhRvL.131r1902G}.
Although both $c_0$ and the gap raise the boundary pressure, they shift the boundary pressure and density in different proportions, so requiring the EOS to match $p$ and $n$ simultaneously can break their degeneracy.
Consequently $c_0$ and $\Delta_{\mathrm{CFL}}^{*}$ remain only mildly correlated, and dropping the CFL term shifts the inferred $c_0$ marginally (to $-19.9^{+10.7}_{-8.9}$).

Third, the non-vanishing peak in the gap posterior, while not yet statistically significant against $\Delta_{\rm CFL}^{*}=0$, hints at a pairing-induced contribution in the matching window.
If corroborated by future advances in pQCD (most notably the as-yet-unknown four-loop hard contribution that fixes $c_0$) and by improved NS observations from NICER and eXTP \cite{2019SCPMA..6229502Z, 2019SCPMA..6229503W, 2025SCPMA..6819503L} of massive pulsars and precise radii, such a signal would constitute the first empirical fingerprint of color superconductivity in dense QCD matter and would directly inform the microscopic mechanism governing CFL pairing.

We note that in our work the N$^3$LO pQCD pressure is evaluated in the massless-quark limit, whereas the strange quark carries a finite mass.
The leading mass correction scales as $\bar m_{\rm s}^2\equiv(3m_{\rm s}/\mu_B)^2=(m_{\rm s}/\mu_q)^2$; at the matching point $\mu_B=2.6$ GeV, this amounts to a fractional shift of only $\sim1-2\%$ in the boundary pressure, far smaller than the renormalization-scale uncertainty ($X\in[1/2,2]$; see Fig.~\ref{fig:mu_p_n}) and the EOS-modeling uncertainty in the intermediate-density window.
The massless N$^3$LO pressure is therefore an adequate approximation for the present purpose.
As explicit cross-checks we repeated the analysis (i) with the mass-corrected NLO pQCD pressure of ref.~\cite{2025PhRvL.135u1901G}, and (ii) with the neural-network bridge in the high-density window $[8\,n_{\rm s},n_{\rm match}]$ replaced by a boundary-constrained wavelet parametrization.
In both cases the inferred $\Delta_{\rm CFL}^{*}$ and the reconstructed EOS posteriors are essentially consistent, confirming that our conclusions are robust against these modeling choices.


\newpage
\begin{center}
{\bf Methods}
\end{center}

Our inference jointly samples (i) the dense-matter EOS represented in sound-speed space, (ii) the pQCD matching parameters, (iii) the CFL pairing parameters, and (iv) auxiliary observational nuisance parameters.
For each sample we build a thermodynamically consistent EOS from the crust up to $100\,n_{\rm s}$, match it at $\mu_B=2.6$ GeV to the N$^3$LO pQCD pressure including the NLO CFL contribution, solve the Tolman-Oppenheimer-Volkoff (TOV) equations \cite{1939PhRv...55..364T,1939PhRv...55..374O}, and evaluate the astrophysical likelihood.
The posterior is obtained by nested sampling.
The control run without color superconductivity is obtained by setting $p_{\rm CFL}^{\rm NLO}=0$ and removing $(\Delta_{\rm CFL}^{*},\sigma)$ from the sampled parameters, while keeping all other settings fixed.

{\bf Hybrid GP--neural-network EOS construction.}
We parameterize the squared sound speed $c_{\rm s}^2(n)$ through the logit variable
\begin{equation}
\phi\equiv-\ln\!\left(1/c_{\rm s}^2-1\right),\qquad c_{\rm s}^2=\sigma(\phi)=\frac{1}{1+e^{-\phi}},
\end{equation}
which enforces $0<c_{\rm s}^2<1$ automatically.
All construction and interpolation are performed in $\phi(\ln n)$ space and evaluated on a fixed grid of $N_{\rm eval}=512$ points logarithmically spaced between $n_{\rm min}\approx 0.33\,n_{\rm s}$ and $100\,n_{\rm s}$.

The EOS is assembled piecewise on the log-density grid from four contiguous segments, joined by a monotone PCHIP interpolation \cite{1980SJNA...17..238F} of $\phi(\ln n)$ that guarantees $C^1$ continuity without overshoots:
(1) {\it Crust and outer core} ($n\lesssim n_{\rm EFT}$): the SLy crust EOS \cite{2001A&A...380..151D} is matched continuously to a $\chi$EFT band.
We sample over the 1000 $\chi$EFT realizations of ref.~\cite{2019PhRvL.122d2501D} (sorted by pressure at each value of $n_{\rm EFT}$) through an integer index $i_{\rm EFT}$, and over the matching density $n_{\rm EFT}/n_{\rm s}\in[1.1,1.8]$.
(2) {\it NS densities} ($n_{\rm EFT}\le n\le 8\,n_{\rm s}$): $\phi$ is described by a GP representation on $N_{\rm bin}=36$ linearly spaced nodes.
(3) {\it Bridge to pQCD} ($8\,n_{\rm s}\le n\le n_{\rm match}$): $\phi$ is described by a boundary-constrained feed-forward neural network (NN).
(4) {\it pQCD region} ($n_{\rm match}\le n\le 100\,n_{\rm s}$): $\phi(n)$ is fixed to the theoretical $c_{\rm s}^2(n)$ from pQCD$+$CFL.
The matching density $n_{\rm match}=n_{\rm pQCD+CFL}(\mu_B=2.6~{\rm GeV})$ is set by the theory itself for the sampled parameters $(\Delta_{\rm CFL}^{*},\sigma,c_0,X)$ and is clipped to $[12,95]\,n_{\rm s}$ for numerical safety.
The left boundary of the bridge inherits the value of the last GP node; its right boundary and the entire pQCD segment are fixed by the theoretical $c_{\rm s}^2(n)$ curve.
An optional Maxwell-type phase-transition module (window parameters $x_{\rm PT}^{\rm lo}$, $x_{\rm PT}^{\rm hi}$ and mixing strength $\alpha_{\rm PT}$, detailed below) softens $c_{\rm s}^2$ inside a smoothly windowed density interval.

We adopt different representations in the two density ranges, $n_{\rm EFT}\le n\le 8\,n_{\rm s}$ and $8\,n_{\rm s}\le n\le n_{\rm match}$, because they pose different demands.
At NS densities the EOS is tightly anchored by $\chi$EFT, so a GP whose mean and variance are calibrated to the $\chi$EFT band provides a physically motivated, well-conditioned, and statistically natural description of the residual uncertainty.
Above $\sim 8\,n_{\rm s}$, by contrast, the EOS must traverse a very wide and weakly constrained density interval up to $n_{\rm match}$ (which typically reaches $\sim 40\,n_{\rm s}$) while smoothly connecting to the pQCD$+$CFL boundary;
covering this range with a GP would demand either many strongly correlated nodes or an ad hoc stiffening of the kernel, inflating the effective dimensionality.
A compact NN with built-in boundary conditions instead spans this bridge with only a handful of parameters, capturing multi-scale structure while keeping the joint sampling tractable.

{\bf GP representation.}
In the NS-density window we model $\bm{\phi}=\{\phi_i\}_{i=1}^{36}$ as a multivariate Gaussian with mean $\bm{\mu}(\bm{\theta}_{\rm GP})$ and covariance $\mathbf{K}(\bm{\theta}_{\rm GP})$ that depend on four sampled hyperparameters $\bm{\theta}_{\rm GP}=(\ell_{\rm GP},\sigma^2_{\rm GP}, \bar c_{\rm s}^2, n_{\rm end})$.
To keep the sampling well conditioned we use a whitening (non-centered) transform \cite{2007StaSc..2200014P},
\begin{equation}
\bm{\phi}=\bm{\mu}(\bm{\theta}_{\rm GP})
          +\mathbf{L}(\bm{\theta}_{\rm GP})\,\bm{u},~
\mathbf{L}\mathbf{L}^{\!\top}=\mathbf{K},~
u_i\sim\mathcal{N}(0,1),
\end{equation}
so that the bin variables are sampled as independent standard normals $u_i$ and mapped to $\phi_i$ via the Cholesky factor $\mathbf{L}$ computed for each draw.

The prior mean interpolates from the $\chi$EFT median toward a conformal-like value $\bar c_{\rm s}^2$ at high density, written as a linear function of $\bar\phi\equiv\mathrm{logit}(\bar c_{\rm s}^2)$, $\mu_i=\mu^{\rm base}_i+w_i\,\bar\phi$.
Defining a pivot density $n_\star=1.8\,n_{\rm s}$ with $\phi_\star=\phi_{\chi\rm EFT}^{\rm med}(n_\star)$ and a sampled smoothstep endpoint $n_{\rm end}$ (with prior centered on $4.55\,n_{\rm s}$), we set: for $n_i\le n_\star$, $\mu^{\rm base}_i=\phi_{\chi\rm EFT}^{\rm med}(n_i)$ and $w_i=0$; for $n_\star<n_i\le n_{\rm end}$, $\mu^{\rm base}_i=(1-s_i)\phi_\star$ and $w_i=s_i$, with $s_i=3t_i^2-2t_i^3$ and $t_i=(n_i-n_\star)/(n_{\rm end}-n_\star)$; for $n_i>n_{\rm end}$, $\mu^{\rm base}_i=0$ and $w_i=1$.
We use a non-stationary RBF kernel
\begin{equation}
K_{ij}=\sigma_i\,\sigma_j\,
\exp\!\left[-\frac{(n_i-n_j)^2}{2\,\ell_{\rm GP}^2}\right]+\epsilon\,\delta_{ij},
\end{equation}
with jitter $\epsilon=10^{-3}$.
The bin-wise standard deviation is fixed by the $\chi$EFT sample variance below $n_\star=1.8\,n_s$, i.e., $\sigma_i^2=\mathrm{Var}[\phi_{\chi\rm EFT}(n_i)]$, and is set to the sampled $\sigma^2_{\rm GP}$ above $n_\star$.
The four hyperparameters carry truncated-Gaussian priors \cite{2023ApJ...950..107G,2024PhRvD.109d3052F},
\begin{align}
\ell_{\rm GP}/n_{\rm s} &\sim \mathcal{T\!N}(0.5,\,0.25;\,[0.1,3.0]),\\
\sigma^2_{\rm GP} &\sim \mathcal{T\!N}(1.25,\,0.20;\,[0.1,3.0]),\\
\bar c_{\rm s}^2 &\sim \mathcal{T\!N}(0.5,\,0.25;\,[0.01,0.99]),\\
n_{\rm end}/n_{\rm s} &\sim \mathcal{T\!N}(4.55,\,1.5;\,[2.2,8.0]),
\end{align}
where $\mathcal{T\!N}(\mu,\sigma;[a,b])$ denotes a Gaussian of mean $\mu$ and width $\sigma$ truncated to $[a,b]$.

{\bf Neural-network bridge.}
Across the wide bridge region $[8\,n_{\rm s},n_{\rm match}]$ we use a boundary-constrained feed-forward NN \cite{2021ApJ...919...11H,2023SciBu..68..913H}.
Let $x\in[0,1]$ be the normalized log-density, $x=(\ln n-\ln 8n_{\rm s})/(\ln n_{\rm match}-\ln 8n_{\rm s})$.
The network output
\begin{equation}
\phi(x)=\underbrace{\phi_{\rm L}+(\phi_{\rm R}-\phi_{\rm L})\,x}_{\text{linear baseline}}
        +\underbrace{x(1-x)}_{\text{envelope}}\,\mathrm{MLP}(x)
\end{equation}
exactly satisfies the boundary conditions $\phi(0)=\phi_{\rm L}$ and $\phi(1)=\phi_{\rm R}$, where $\phi_{\rm L}$ is the value of the last GP node and $\phi_{\rm R}=\mathrm{logit}\,c_{\rm s}^2(n_{\rm match})$ is set by the pQCD$+$CFL theory.
The $\mathrm{MLP}$ is a single hidden layer of $10$ neurons with $\tanh$ activation,
\begin{equation}
\mathrm{MLP}(x)=\bm{W}_{\rm out}\tanh(\bm{W}_1 x+\bm{b}_1)+b_{\rm out}.
\end{equation}
The hidden-layer weights carry Xavier-scaled Gaussian priors $\bm{W}_1\sim\mathcal{N}(0,(\kappa\,\sigma_{\rm X})^2)$, with $\sigma_{\rm X}=\sqrt{2/(n_{\rm in}+n_{\rm out})}=\sqrt{2/11}$ and $\kappa=3$.
To regularize the bridge toward smooth, well-ordered shapes across the wide intermediate window, we adopt shape priors on the remaining parameters: the output-layer weights are non-negative, $\bm{W}_{\rm out}\sim\mathcal{HN}(\kappa\sqrt{2/11})$, and the hidden-layer biases are monotonically ordered through a cumulative construction, $b_1^{(0)}\sim\mathcal{N}(-1.6,1)$ and $b_1^{(i)}=b_1^{(i-1)}+\delta_i$ with positive increments $\delta_i\sim\mathcal{HN}(0.45)$; the output bias uses $b_{\rm out}\sim\mathcal{N}(0,0.5^2)$.
Here $\mathcal{HN}(\sigma)$ denotes a half-Gaussian (one-sided) distribution with scale $\sigma$.
The network is evaluated on $\mathcal{O}(20)$ interior anchor points that are then fed to the global PCHIP interpolation.
This construction preserves the flexibility of nonparametric models while introducing only $31$ parameters across the bridge, keeping the dimensionality tractable for nested sampling.

{\bf Phase-transition module.}
To allow for a (quasi-)first-order softening we optionally superimpose a Maxwell-type window on $\phi(\ln n)$.
Its location is set by two dimensionless parameters $x_{\rm PT}^{\rm lo},x_{\rm PT}^{\rm hi}\in[0,1]$ that map linearly in $\ln n$ onto the interval $[2\,n_{\rm s},\,n_{\rm match}]$,
\begin{equation}
\ln n_{\rm PT}^{\rm lo,hi}=\ln(2\,n_{\rm s})
+x_{\rm PT}^{\rm lo,hi}\,\big[\ln n_{\rm match}-\ln(2\,n_{\rm s})\big],
\end{equation}
subject to the width constraint $x_{\rm PT}^{\rm hi}-x_{\rm PT}^{\rm lo}\in[0,1/2]$ (a positive window no wider than half the available log-density range).
Inside the window, $\phi$ is blended toward the soft value $\phi_{\rm PT}=\mathrm{logit}(10^{-3})$,
\begin{equation}
\phi(\ln n)\;\longrightarrow\;
\big[1-\alpha_{\rm PT}\,W(\ln n)\big]\,\phi(\ln n)
+\alpha_{\rm PT}\,W(\ln n)\,\phi_{\rm PT},
\end{equation}
where $\alpha_{\rm PT}\in[0,1]$ is the mixing strength and
\begin{equation}
W(\ln n)=\varsigma\!\big[s(\ln n-\ln n_{\rm PT}^{\rm lo})\big]\;
         \varsigma\!\big[-s(\ln n-\ln n_{\rm PT}^{\rm hi})\big]
\end{equation}
is a smooth top-hat built from logistic functions $\varsigma(z)=(1+e^{-z})^{-1}$ with edge sharpness $s=12$.
Thus $\alpha_{\rm PT}=0$ recovers the smooth EOS, whereas $\alpha_{\rm PT}\to 1$ drives $c_{\rm s}^2$ toward $\sim10^{-3}$ across the window, producing a near-plateau in $p(\epsilon)$.
As shown in the Supplementary Information, the posterior pushes $(n_{\rm PT}^{\rm lo},n_{\rm PT}^{\rm hi})$ to densities around and above $n_{\rm TOV}$, leaving the transition largely inactive within the bulk of the stable NS sequence.

{\bf pQCD pressure at N$^3$LO and the CFL gap.}
We adopt the massless N$^3$LO pQCD pressure of ref.~\cite{2023PhRvL.131r1902G},
\begin{align}
\frac{p_{\rm pQCD}^{\rm N^3LO}}{p_{\rm free}}=
1&-\frac{2\alpha_{\rm s}}{\pi}-3\!\left(\frac{\alpha_{\rm s}}{\pi}\right)^{\!2}\!\Big[L+3\ln X+5.0021\Big]
\nonumber\\
&+9\!\left(\frac{\alpha_{\rm s}}{\pi}\right)^{\!3}\!\Big[\tfrac{11}{12}L^2+(-6.5968-3\ln X)L
\nonumber\\
&\quad+\big(5.1342+\tfrac{2}{3}c_0-18.284\ln X-4.5\ln^2 X\big)\Big],
\end{align}
where $p_{\rm free}=\mu_B^4/(108\pi^2)$, $L\equiv\ln(3\alpha_{\rm s}/\pi)$, and $X\equiv 3\bar\Lambda/(2\mu_B)\in[0.5,2]$ is the renormalization-scale ratio.
The single unknown $c_0$ encodes the IR-finite remainder of the unresummed four-loop diagrams.
The CFL color-superconducting contribution is taken at NLO under NS conditions \cite{2025PhRvL.135u1901G},
\begin{equation}
p_{\rm CFL}^{\rm NLO}=p_{\rm free}\!\left[\,\gamma_1(\alpha_s,\bar m_s^2)\,
\bar{\Delta}_{\rm CFL}^2-\frac{\bar m_s^4}{4}\,\right],
\end{equation}
with
\begin{equation}
\gamma_1=4-\frac{4}{3}\bar m_{\rm s}^2+40.9\,\alpha_{\rm s},~
\bar{\Delta}_{\rm CFL}=\frac{\Delta_{\rm CFL}}{\mu_B/3},~
\bar m_{\rm s}=\frac{m_{\rm s}}{\mu_B/3}.
\end{equation}
We allow a power-law running of the gap, $\Delta_{\rm CFL}(\mu_B)=\Delta_{\rm CFL}^{*}(\mu_B/\mu_B^{*})^{\sigma}$ with $\mu_B^{*}=2.6$ GeV.
The baryon density and sound speed used to set $n_{\rm match}$ and the pQCD segment of $c_{\rm s}^2(n)$ are obtained analytically from the first and second $\mu_B$-derivatives of $p_{\rm pQCD}^{\rm N^3LO}+p_{\rm CFL}^{\rm NLO}$, including the running of $\alpha_{\rm s}$, $m_{\rm s}$, and of the prefactor $\gamma_1$.
The running coupling $\alpha_{\rm s}(\mu)$ and strange-quark mass $m_s(\mu)$ are obtained by integrating the four-loop $\overline{\rm MS}$ $\beta$ and $\gamma_m$ functions from $\alpha_{\rm s}(2~{\rm GeV})=0.2994$ and $m_s(2~{\rm GeV})=93.8$ MeV \cite{2008PhLB..667....1A, 2025PhRvL.135u1901G}, evaluated at the renormalization scale $\bar\Lambda=2X\mu_B/3$.

{\bf Matching condition.}
For a given draw with sampled $(\Delta_{\rm CFL}^{*},\sigma,c_0,X)$, the pQCD$+$CFL theory provides two target values at the boundary $\mu_B^{*}=2.6$ GeV:
\begin{align}
p_{\rm th} &\equiv p_{\rm pQCD}^{\rm N^3LO}(\mu_B^{*};X,c_0)
                  +p_{\rm CFL}^{\rm NLO}(\mu_B^{*};X,\Delta_{\rm CFL}^{*},\sigma),\\
n_{\rm th} &\equiv \left.\frac{\partial p}{\partial\mu_B}\right|_{\mu_B^{*}},
\end{align}
where $n_{\rm th}$ is computed analytically from the same pressure (including the running of $\alpha_{\rm s}$, $m_{\rm s}$, and $\gamma_1$).
The $n_{\rm th}$ fixes the matching density $n_{\rm match}$ used in the EOS construction.
On the EOS side, we integrate the thermodynamic relations to obtain $p(\mu_B)$ and $n(\mu_B)$, and read off the values at the boundary by interpolation, $p\equiv p(\mu_B^{*})$ and $n\equiv n(\mu_B^{*})$.
Consistency between the parameterized EOS and the theory is imposed through two Gaussian penalties added to the log-likelihood,
\begin{equation}
\ln\mathcal{L}_{\rm match}=
-\frac{(n-n_{\rm th})^2}{2\,(\sigma_n\,n_{\rm th})^2}
-\frac{(p-p_{\rm th})^2}{2\,(\sigma_p\,p_{\rm th})^2},
\end{equation}
with relative density and pressure tolerances $\sigma_n=\sigma_p=0.01$.

{\bf Observational likelihood and sampling.}
The astrophysical likelihood combines:
(i) the posterior on $M_{\rm max}$ from the population analysis of all rotation-corrected NS masses \cite{2024PhRvD.109d3052F}, entering as a soft maximum-mass constraint;
(ii) the GW170817 tidal-deformability information via a random-forest surrogate of the marginal GW likelihood \cite{2020MNRAS.499.5972H}, built upon the IMRPhenomPv2NRT low-spin analysis \cite{2019PhRvX...9a1001A}, with the binary chirp mass and mass ratio sampled jointly;
(iii) NICER mass-radius posteriors for PSR J0030$+$0451 (ST$+$PDT) \cite{2024ApJ...961...62V, 2024ApJ...966...98L}, PSR J0740$+$6620 \cite{2024ApJ...974..294S}, PSR J0437$-$4715 \cite{2024ApJ...971L..20C}, and PSR J0614$-$3329 \cite{2025ApJ...995...60M}, evaluated through kernel-density estimates.
For the N$^3$LO constant we use the weakly informed prior on $c_0$ derived from the convergence analysis of the N$^3$LO pressure \cite{2023PhRvL.131r1902G}, restricted to $c_0\in[-40,10]$; the CFL gap magnitude $\Delta_{\rm CFL}^{*}\in[0,300]$ MeV and exponent $\sigma\in[-0.5,0.5]$ carry uniform priors, and $X$ is sampled log-uniformly in $[0.5,2]$.
Posteriors are obtained with \texttt{dynesty} nested sampling \cite{2020MNRAS.493.3132S} using $N_{\rm live}=5000$ live points and \texttt{rwalk} moves, evolving the full set of EOS, GW, NICER, pQCD, and CFL parameters jointly.

\begin{addendum}
\item[Acknowledgements] This work is supported by the National Natural Science Foundation of China under Grants No. 12588101, No. 12233011, and No. 12303056, the Project for Special Research Assistant and the Project for Young Scientists in Basic Research (No. YSBR-088) of the Chinese Academy of Sciences, and the Postdoctoral Fellowship Program of China Postdoctoral Science Foundation (GZC20241915).

\item[Author Contributions] Y.Z.F. conceived the project. S.P.T. conducted the coding and data analysis and wrote the initial draft. All authors discussed the results and revised the manuscript.

\item[Author Information] Correspondence and requests for materials should be addressed to Y.Z.F.~(yzfan@pmo.ac.cn).
\end{addendum}

\newpage
\renewcommand{\refname}{References}
\bibliographystyle{sn-nature}
\bibliography{refs.bib}


\newpage

\begin{center}
{\bf Supplementary Information}
\end{center}

This Supplementary Information collects the supplemental posterior distributions referenced in the main text.
Unless stated otherwise the quoted numbers refer to the fiducial N$^3$LO pQCD$+$CFL inference; for the corner plots we also overlay the control run without the CFL contribution (w/o CFL).

\paragraph{Mass-radius posterior.}
Supplementary Figure~\ref{fig:mr} shows the posterior mass-radius bands (90\% CI) for the inferences with and without the CFL contribution, overlaid with the observational inputs: the GW170817 primary and secondary components, and the NICER measurements of PSR~J0030$+$0451, PSR~J0740$+$6620, PSR~J0437$-$4715, and PSR~J0614$-$3329.
The inference supports $M_{\rm TOV}\simeq 2.16\,M_\odot$ and a moderately compact radius $R_{1.4}\simeq 11.5$ km, comfortably accommodating all current constraints, and the two bands are nearly coincident.

\paragraph{Phase-transition and matching parameters.}
Supplementary Figure~\ref{fig:corner_corr} shows the joint posterior of the phase-transition and matching parameters ($n_{\rm PT}^{\rm lo}$, $n_{\rm PT}^{\rm hi}$, $\alpha_{\rm PT}$, $c_0$, $X$, $\Delta_{\rm CFL}^{*}$, $\sigma$).
The Maxwell-type phase-transition window is broad and centered at high density ($n_{\rm PT}^{\rm lo}\simeq 5.8\,n_{\rm s}$, $n_{\rm PT}^{\rm hi}\simeq 13.1\,n_{\rm s}$, with mixing strength $\alpha_{\rm PT}\simeq 0.40$), i.e., around and above $n_{\rm TOV}$, so that it has little effect on the bulk of the stable NS sequence.
The renormalization-scale ratio settles at $X\simeq 0.92$, and $c_0$ and $\Delta_{\rm CFL}^{*}$ show only mild correlation.

\paragraph{Derived bulk NS properties.}
Supplementary Figure~\ref{fig:meta} shows the corner plot of the derived bulk NS properties, including $M_{\rm TOV}$, $R_{\rm TOV}$, $R_{1.4}$, $R_{2.0}$, $\Lambda_{1.4}$, the $\chi$EFT--GP matching density $n_{\rm EFT}$, and the central density $n_{\rm TOV}$ of the maximum-mass star.
The corresponding quantitative constraints (w/ CFL) are
$M_{\rm TOV}=2.16^{+0.10}_{-0.09}\,M_\odot$,
$R_{\rm TOV}=11.06^{+0.60}_{-0.53}$ km,
$R_{1.4}=11.49^{+0.56}_{-0.43}$ km,
$R_{2.0}=11.37^{+0.56}_{-0.49}$ km,
$\Lambda_{1.4}=250^{+98}_{-56}$,
$n_{\rm EFT}=1.40^{+0.32}_{-0.25}\,n_{\rm s}$, and
$n_{\rm TOV}=6.40^{+0.70}_{-0.64}\,n_{\rm s}$.
The w/o CFL run gives statistically indistinguishable values and is shown for comparison in the same figure.

\paragraph{GP hyperparameters.}
Supplementary Figure~\ref{fig:hyper} shows the posterior of the four GP hyperparameters ($\ell_{\rm GP}$, $\sigma^2_{\rm GP}$, $\bar c_{\rm s}^2$, $n_{\rm end}$).
We find $\ell_{\rm GP}=0.52^{+0.26}_{-0.21}\,n_{\rm s}$, $\sigma^2_{\rm GP}=1.27^{+0.25}_{-0.22}$, $\bar c_{\rm s}^2=0.53^{+0.20}_{-0.19}$, and $n_{\rm end}=3.90^{+1.05}_{-0.98}\,n_{\rm s}$.
The data therefore prefer a moderate correlation length and a sizable amplitude, allowing structure such as the $c_{\rm s}^2$ peak while remaining smooth.
The prior-mean transition density $n_{\rm end}$ settles near the peak location, and the asymptotic sound speed $\bar c_{\rm s}^2$ remains close to $0.5$, in agreement with the bands in Fig.~\ref{fig:eos}.

\renewcommand{\figurename}{Supplementary Figure}
\setcounter{figure}{0}

\begin{figure}[!h]
\centering
\includegraphics[width=0.618\linewidth]{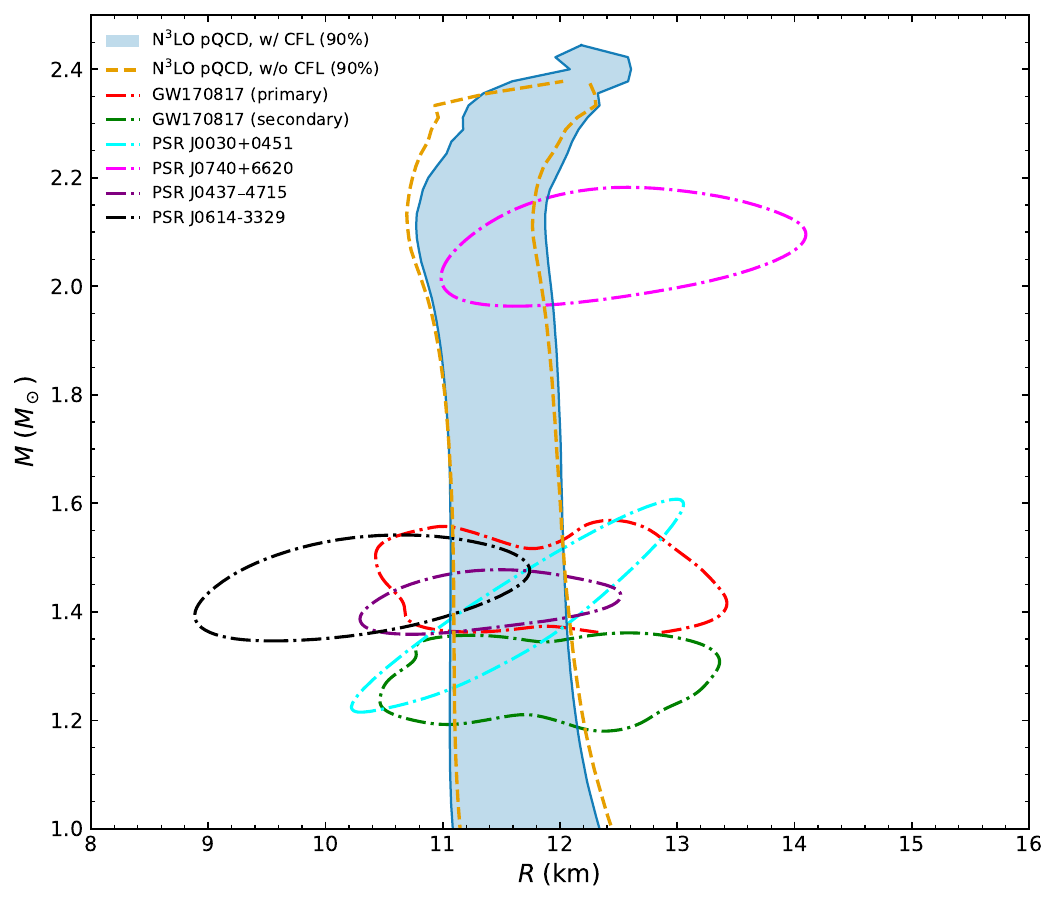}
\caption{Posterior mass-radius bands (90\% CI) for the inferences with (w/ CFL) and without (w/o CFL) the color-superconducting contribution, compared with the mass-radius measurements for PSR J0030+0451, PSR J0740+6620, PSR J0437--4715, PSR J0614--3329, and the components of the GW170817 event (68.3\% CI).}
\label{fig:mr}
\end{figure}

\begin{figure}
\centering
\includegraphics[width=0.96\linewidth]{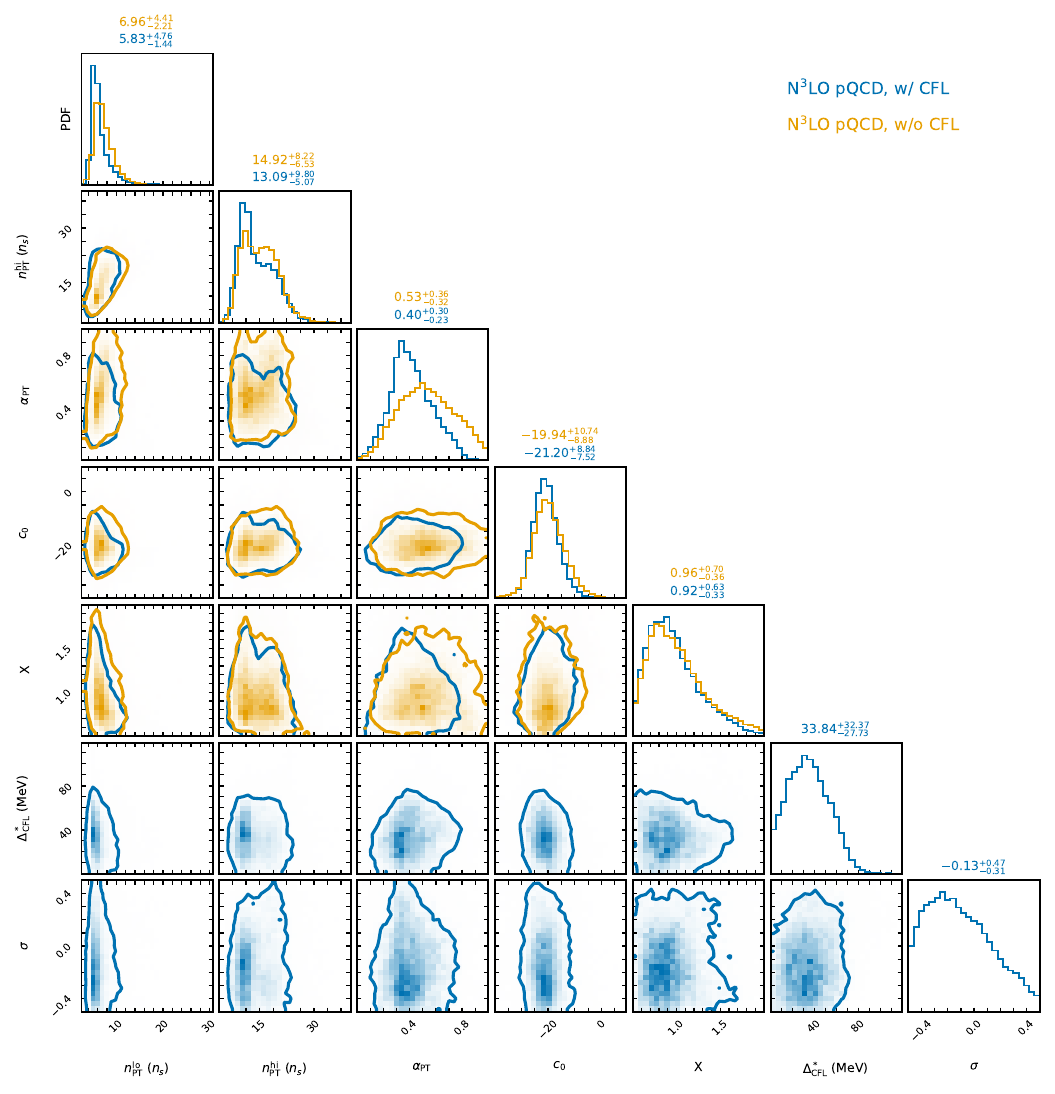}
\caption{Corner plot of the phase-transition and matching parameters ($n_{\rm PT}^{\rm lo}$, $n_{\rm PT}^{\rm hi}$, $\alpha_{\rm PT}$, $c_0$, $X$, $\Delta_{\rm CFL}^{*}$, $\sigma$) for the inferences with (w/ CFL) and without (w/o CFL) the pairing term. The phase-transition densities lie around and above $n_{\rm TOV}$, and the pQCD constant is constrained to $c_0\simeq-21$.}
\label{fig:corner_corr}
\end{figure}

\begin{figure}
\centering
\includegraphics[width=0.96\linewidth]{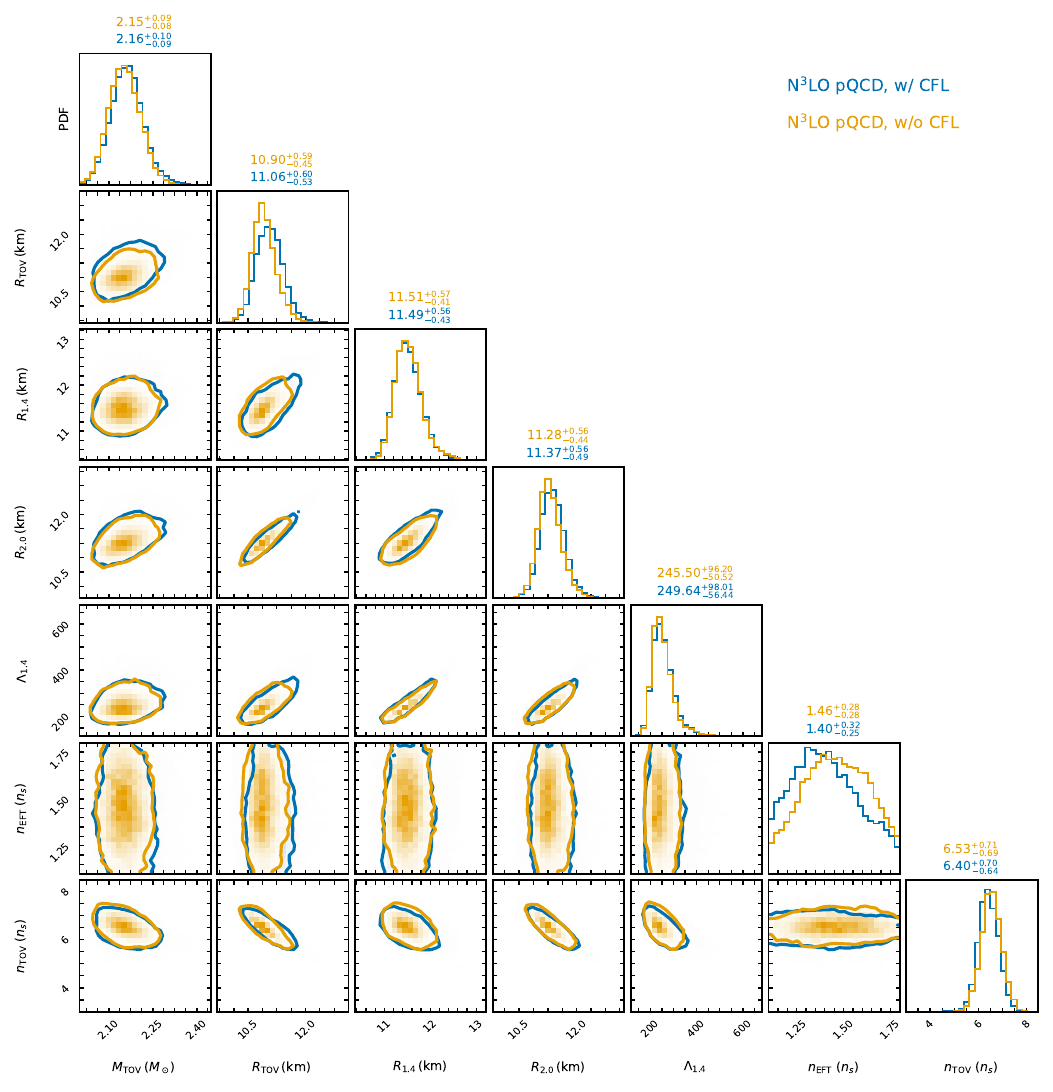}
\caption{Corner plot of the derived macroscopic NS observables, the $\chi$EFT$-$GP connecting density $n_{\rm EFT}$, and the central density $n_{\rm TOV}$ of the maximum-mass star, for the inferences with (w/ CFL) and without (w/o CFL) the color-superconducting contribution.}
\label{fig:meta}
\end{figure}

\begin{figure}[!h]
\centering
\includegraphics[width=0.618\linewidth]{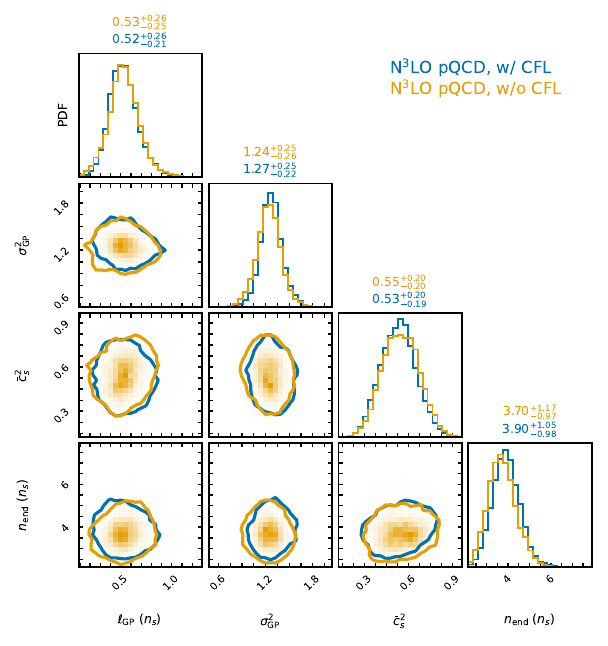}
\caption{Posterior of the GP hyperparameters $\ell_{\rm GP}$, $\sigma^2_{\rm GP}$, $\bar c_{\rm s}^2$, and $n_{\rm end}$, for the inferences with (w/ CFL) and without (w/o CFL) the color-superconducting contribution.}
\label{fig:hyper}
\end{figure}

\end{document}